\documentclass[aps,prl,reprint,superscriptaddress,floatfix,letterpaper]{revtex4-1}
\usepackage[utf8]{inputenc}

\newcommand{\bra}[1]{\langle #1|}
\newcommand{\ket}[1]{|#1\rangle}
\newcommand{\avg}[1]{\langle #1 \rangle}

\newcommand{\tr}{\mathrm{Tr}}

\usepackage{amsmath}
\usepackage{amssymb}

\usepackage{graphicx}

\usepackage[urlcolor=blue, hyperindex, colorlinks, bookmarks=true]{hyperref}

\begin{document}

\title{Quantum Non-demolition Detection of Single Microwave Photons in a Circuit}

\author{B. R. Johnson}
\author{M. D. Reed}
\affiliation{Departments of Physics and Applied Physics, Yale University, New Haven, CT 06511, USA}
\author{A. A. Houck}
\affiliation{Department of Electrical Engineering, Princeton University, Princeton, NJ 08544, USA}
\author{D. I. Schuster}
\author{Lev S. Bishop}
\author{E. Ginossar}
\affiliation{Departments of Physics and Applied Physics, Yale University, New Haven, CT 06511, USA}
\author{J. M. Gambetta}
\affiliation{Institute for Quantum Computing and Department of Physics and Astronomy, University of Waterloo, Waterloo, ON, Canada, N2L 3G1}
\author{L. DiCarlo}
\author{L. Frunzio}
\author{S. M. Girvin}
\author{R. J. Schoelkopf}
\affiliation{Departments of Physics and Applied Physics, Yale University, New Haven, CT 06511, USA}

\date{March 12, 2010}

\ifpdf
\DeclareGraphicsExtensions{.pdf, .jpg, .tif}
\else
\DeclareGraphicsExtensions{.eps, .jpg}
\fi

\maketitle

\textbf{Thorough control of quantum measurement is key to the development of quantum information technologies. Many measurements are destructive, removing more information from the system than they obtain. Quantum non-demolition (QND) measurements allow repeated measurements that give the same eigenvalue~\cite{Braginsky:1996}. They could be used for several quantum information processing tasks such as error correction~\cite{Steane:1996}, preparation by measurement~\cite{Ruskov:2003}, and one-way quantum computing~\cite{Raussendorf:2001}. Achieving QND measurements of photons is especially challenging because the detector must be completely transparent to the photons while still acquiring information about them~\cite{Guerlin:2007, Gambetta:2006}. Recent progress in manipulating microwave photons in superconducting circuits~\cite{Houck:2007, Hofheinz:2008, Hofheinz:2009} has increased demand for a QND detector which operates in the gigahertz frequency range. Here we demonstrate a QND detection scheme which measures the number of photons inside a high quality-factor microwave cavity on a chip. This scheme maps a photon number onto a qubit state in a single-shot via qubit-photon logic gates. We verify the operation of the device by analyzing the average correlations of repeated measurements, and show that it is 90\% QND\@. It differs from previously reported detectors~\cite{Brune:1996, Schuster:2007, Hofheinz:2008, Hofheinz:2009, Guerlin:2007} because its sensitivity is strongly selective to chosen photon number states. This scheme could be used to monitor the state of a photon-based memory in a quantum computer.}

Several teams have engineered detectors which are sensitive to single microwave photons by strongly coupling atoms (or artificial atoms) to high-Q cavities. This architecture, known as cavity quantum electrodynamics (cavity QED), can be used in various ways to detect photons. One destructive method measures quantum Rabi oscillations of an atom or qubit \textit{resonantly} coupled to the cavity~\cite{Brune:1996, Hofheinz:2008, Hofheinz:2009}. The oscillation frequency is proportional to $\sqrt{n}$, where $n$ is the number of photons in the cavity, so this method essentially measures the time-domain swap frequency.

Another method uses a \textit{dispersive} interaction to map the photon number in the cavity onto the phase difference of a superposition of atomic states $(\ket{g} + e^{i\phi}\ket{e})/\sqrt{2}$. Each photon number $n$ corresponds to a different phase $\phi$, so repeated Ramsey experiments~\cite{Guerlin:2007} can be used to estimate the phase and extract $n$. This method is QND, because it does not exchange energy between the atom and photon. However, since the phase cannot be measured in a single operation, it does not extract full information about a particular Fock state $\ket{n}$ in a single interrogation. Nonetheless, using Rydberg atoms in cavity QED, remarkable experiments have shown quantum jumps of light and the collapse of the photon number by measurement.~\cite{Guerlin:2007,Gleyzes:2007}

Here we report a new method which implements a set of programmable controlled-NOT (CNOT) operations between an $n$-photon Fock state and a qubit, asking the question ``are there exactly $n$ photons in the cavity?" A single interrogation consists of applying one such CNOT operation and reading-out the resulting qubit state. To do this we use a quasi-dispersive qubit-photon interaction which causes the qubit transition frequency to depend strongly on the number of photons in the cavity. Consequently, frequency control of a pulse implements a conditional $\pi$ rotation on the qubit -- the qubit state is inverted if and only if there are $n$ photons in the storage cavity. To ensure that this is QND, the qubit and storage cavity are adiabatically decoupled before performing a measurement of the qubit state.

\begin{figure}
	\includegraphics{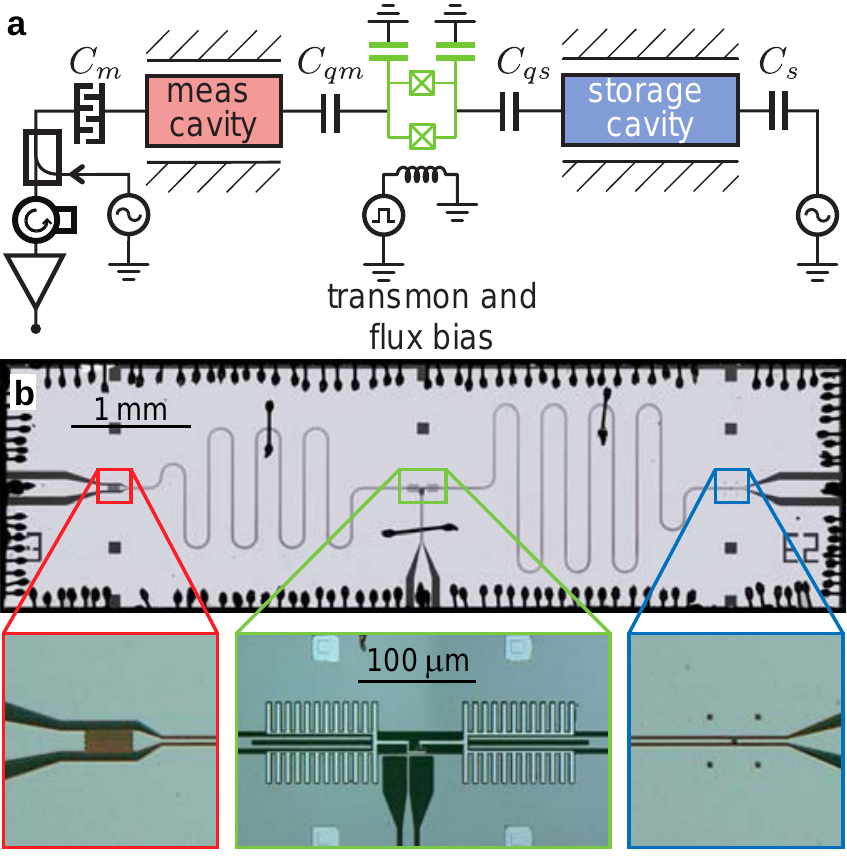}
	\caption{\label{figure 1}Circuit schematic and device. \textbf{a}, Circuit schematic showing two cavities coupled to a single transmon qubit. The measurement cavity is probed in reflection by sending microwave signals through the weakly coupled port of a directional coupler. A flux bias line allows for tuning of the qubit frequency on nanosecond timescales. \textbf{b}, Implementation on a chip, with $\omega_m/2\pi = 6.65\,\mathrm{GHz}$ measurement cavity on the left and its large coupling capacitor (red), and $\omega_s/2\pi = 5.07\,\mathrm{GHz}$ storage cavity on the right with a much smaller coupling capacitor (blue). A transmon qubit (green) is strongly coupled to each cavity, with $g_s/2\pi = 70\,\mathrm{MHz}$ and $g_m/2\pi = 83\,\mathrm{MHz}$. It has a charging energy $E_C/2\pi = 290\,\mathrm{MHz}$ and maximal Josephson energy $E_J/2\pi \approx 23\,\mathrm{GHz}$. At large detunings from both cavities, the qubit coherence times are $T_1 \approx T_2 \approx 0.7\,\mathrm{\mu s}$.}
\end{figure}

To realize this method we extend circuit-based cavity QED~\cite{Schoelkopf:2008} by coupling a single transmon qubit~\cite{Koch:2007, Schreier:2008} simultaneously to two cavities. This allows one cavity to be optimized for fast readout and the other for coherent storage of photons. Related work by Leek \emph{et al.} ~\cite{Leek:2010} realized a single transmon coupled to two modes of a single cavity, where the the two modes were engineered to have very different quality factors. A schematic of the two-cavity device is shown in Fig.\,\ref{figure 1}(a). A high-Q cavity serves as a photon memory for preparation and storage, and a low-Q cavity is used for fast readout of the qubit. The cavities are realized as Nb coplanar waveguide resonators with $\lambda/2$ resonances at $\omega_s/2\pi = 5.07\,\mathrm{GHz}$ and $\omega_m/2\pi = 6.65\,\mathrm{GHz}$, respectively. The cavities are engineered, by design of the capacitors $C_s$ and $C_m$, to have very different decay rates ($\kappa_s/2\pi = 50\,\mathrm{kHz}$ and $\kappa_m/2\pi = 20\,\mathrm{MHz}$) so that the qubit state can be measured several times per photon lifetime in the storage cavity. A transmon qubit is end-coupled to the two cavities, with finger capacitors controlling the individual coupling strengths ($g_s/2\pi = 70\,\mathrm{MHz}$ and $g_m/2\pi = 83\,\mathrm{MHz}$). The usual shunt capacitor between the transmon islands is replaced with capacitors to the ground planes to reduce direct coupling between the cavities. Additionally, a flux bias line~\cite{Dicarlo:2009} allows fast control of the detunings $\Delta_s = \omega_{g,e} - \omega_s$ and $\Delta_m = \omega_{g,e} - \omega_m$ between the transmon and cavities, where we use the convention of labeling the transmon states from lowest to highest energy as ($g$, $e$, $f$, $h$, ...).

\begin{figure}
	\includegraphics{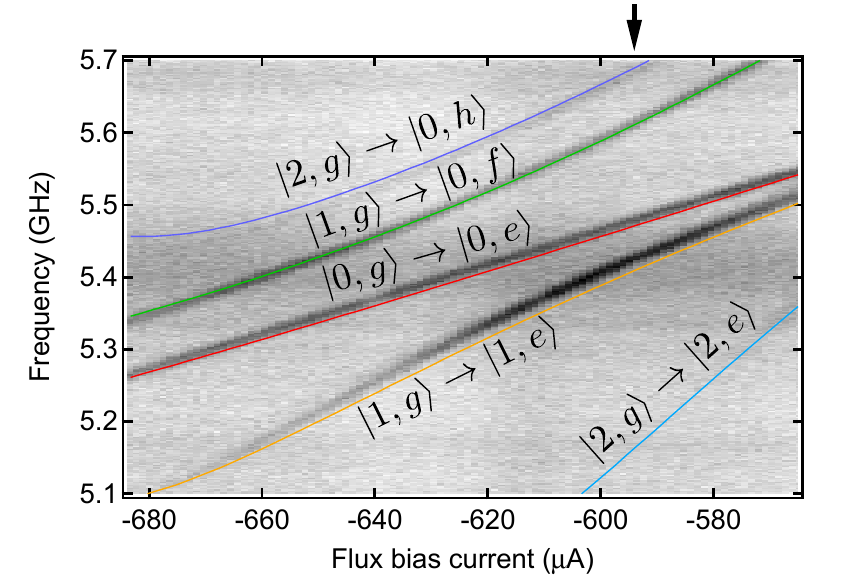}
	\caption{\label{figure 2} Pulsed spectroscopy with coherent state in storage cavity ($\langle n \rangle \approx 1$) vs. qubit-cavity detuning $\Delta_s = \omega_{g,e} - \omega_s$. Calculated transition frequencies are overlaid in color. Red and orange lines are the $\ket{g} \leftrightarrow \ket{e}$ transitions of the qubit when $n=0$ and 1, respectively. Transitions to higher transmon levels ($\ket{f}$ and $\ket{h}$) are visible because of the small detuning. The arrow indicates the flux bias current used during the CNOT operations.}
\end{figure}

To achieve high photon number selectivity of the CNOT operations, there must be a large separation between the number-dependent qubit transition frequencies. To obtain this, we use small detunings ($\Delta_s/g_s < 10$) between the qubit and storage cavity. Figure~\ref{figure 2} shows spectroscopy in this quasi-dispersive regime as a function of flux bias when the storage cavity is populated with a coherent state ($\avg{\hat{n}} \sim 1$). Results of a numerical energy-level calculation are overlaid, showing the positions of various transitions. We define $\omega_{g,e}^{n}$ as the photon number-dependent transition frequency $\ket{n,g} \rightarrow \ket{n,e}$. Other transitions, such as $\ket{2,g} \rightarrow \ket{0,h}$, are allowed due to the small detuning. Fortunately, we also see that the separation between $\omega_{g,e}^{0}$ and $\omega_{g,e}^{1}$ grows rapidly to order $\sim 2g = 140\,\mathrm{MHz}$ as the qubit approaches the storage cavity.

To test the photon meter, we generate single photons in the storage cavity with an adiabatic protocol. Our method uses the avoided crossing between the $\ket{0,e}$ and $\ket{1,g}$ levels to convert a qubit excitation into a photon. The preparation of a photon begins with the qubit detuned below the storage cavity ($\Delta_s \simeq -3 g_s$), where we apply a $\pi$-pulse to create the state $\ket{0,e}$. We then adiabatically tune the qubit frequency through the avoided crossing with the storage cavity, leaving the system in the state $\ket{1,g}$. The sweep rate is limited by Landau-Zener transitions which keep the system in $\ket{0,e}$. Our preparation protocol changes the qubit frequency by 600~MHz in 50~ns, giving a transition probability less than 0.1\% (calculated with a multi-level numerical simulation). This protocol actually allows for the creation of arbitrary superpositions of $\ket{0,g}$ and $\ket{1,g}$ by changing the rotation angle of the initial pulse. For example, if we use a $\pi/2$-pulse, after the sweep we end up in the state $(\ket{0,g} + e^{i\phi}\ket{1,g})/\sqrt{2}$, where $\phi$ is determined by the rotation axis of the $\pi/2$-pulse. One could also use a resonant swap scheme, which has been successfully used to create Fock states~\cite{Hofheinz:2009} up to $\ket{n=15}$. The method used here has the advantage of being very robust to timing errors.

\begin{figure}
	\includegraphics{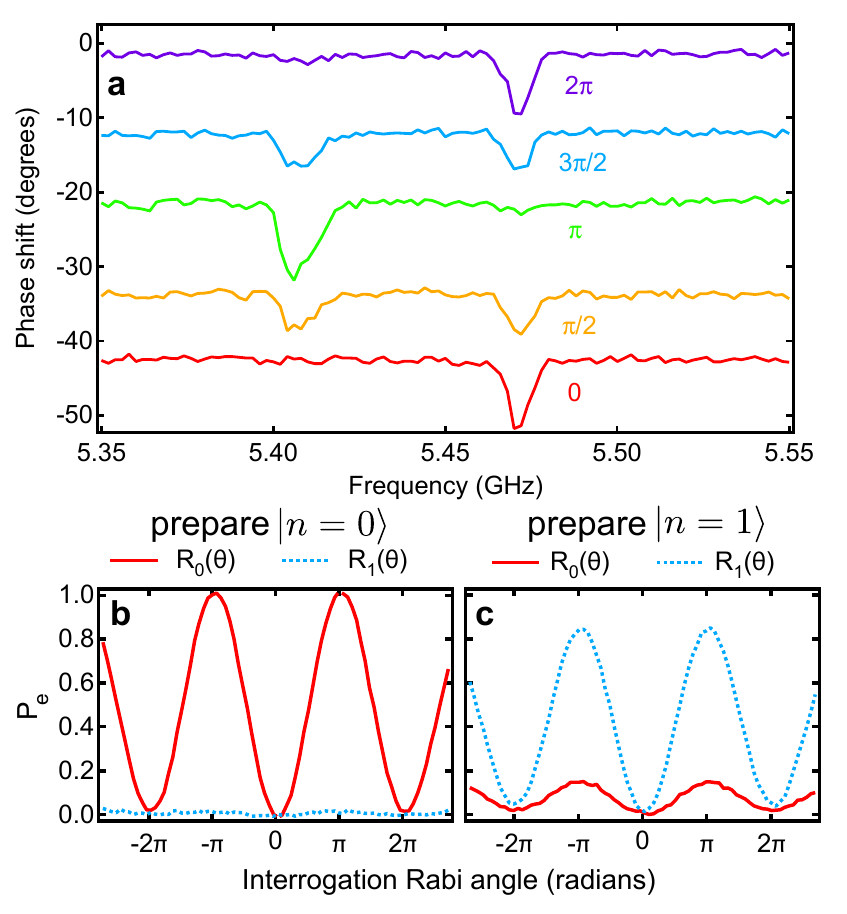}
	\caption{\label{figure 3}Single photon preparation and CNOT selectivity. \textbf{a}, Pulsed spectroscopy vs. Rabi angle of preparation pulse, showing the reflected phase of a pulse at the measurement cavity frequency after a $\sim 80\,\mathrm{ns}$ pulse near the qubit frequency. Traces are offset vertically for clarity and labeled with the rotation angle of the control pulse used in the preparation step. The dips correspond to $\omega_{g,e}^{0} \approx 5.47\,\mathrm{GHz}$ and $\omega_{g,e}^{1} \approx 5.41\,\mathrm{GHz}$, respectively. \textbf{b} and \textbf{c}, Rabi driving the qubit transitions after preparing $\ket{n=0}$ (\textbf{b}) and $\ket{n=1}$ (\textbf{c}). The red (blue) traces show the measured qubit excited state probability after applying an interrogation Rabi pulse with varying angle at $\omega_{g,e}^{0}$ ($\omega_{g,e}^{1}$). The residual oscillation of $R_1(\theta)$ in \textbf{c} is mostly due to preparation infidelity.}
\end{figure}

After the photon is prepared, the qubit frequency is adjusted such that $\Delta_s/g_s \simeq 5$. At this detuning,  the separation between $\omega_{g,e}^{0}$ and $\omega_{g,e}^{1}$ is $\sim 65\,\mathrm{MHz}$. In Fig.\,\ref{figure 3}(a), we show pulsed spectroscopy at this detuning for several rotation angles of the initial preparation pulse. We observe well-resolved dips in the reflected phase of a pulsed signal sent at the measurement cavity frequency. The locations of these dips correspond to the qubit transition frequencies for $n=0$ ($\omega_{g,e}^{0}$) and $n=1$ ($\omega_{g,e}^{1}$), and the relative heights match expectations from the different preparation pulse rotations (e.g. a $\pi/2$-pulse results in equal height signals).

To show selective driving of these transitions, we perform Rabi experiments at $\omega_{g,e}^{0}$ and $\omega_{g,e}^{1}$ for the cases where we prepare $\ket{0,g}$ and $\ket{1,g}$. In each experiment we ensemble average measurements of the resulting qubit state after further decoupling the qubit from the storage cavity. For the $\ket{0,g}$ case [Fig.\,\ref{figure 3}(b)] there is a large amplitude oscillation when the drive is at $\omega_{g,e}^{0}$ [red, $R_0(\theta)$] and almost no oscillation when the drive is at $\omega_{g,e}^{1}$ [blue, $R_1(\theta)$]. When we prepare $\ket{1,g}$ the situation is reversed [Fig.\,\ref{figure 3}(c)]; however, in this case the residual oscillation of $R_0(\theta)$ (red) is substantial due to small errors in the preparation of $\ket{1,g}$ associated with the initial rotation of the qubit and, more importantly, the $\sim 10\%$ probability of energy decay during the subsequent adiabatic sweep through the cavity.

The responses $R_i(\theta)$ are a result of driving $\omega_{g,e}^i$ and the far off-resonant drive of $\omega_{g,e}^j$, where $j\neq i$. The cross-talk is seen in the small residual oscillation of $R_1(\theta)$ in Fig.\,\ref{figure 3}(b). In the supplement, we derive a method for extracting a selectivity and preparation fidelity from these data, giving a selectivity $\geq 95\%$ for both interrogations and a preparation fidelity of $|\langle n=1|\psi\rangle|^2\approx 88\%$. These numbers were confirmed by doing equivalent experiments over a range of preparation pulse rotation angles between 0 and $2\pi$ (not shown).

\begin{figure}
	\includegraphics{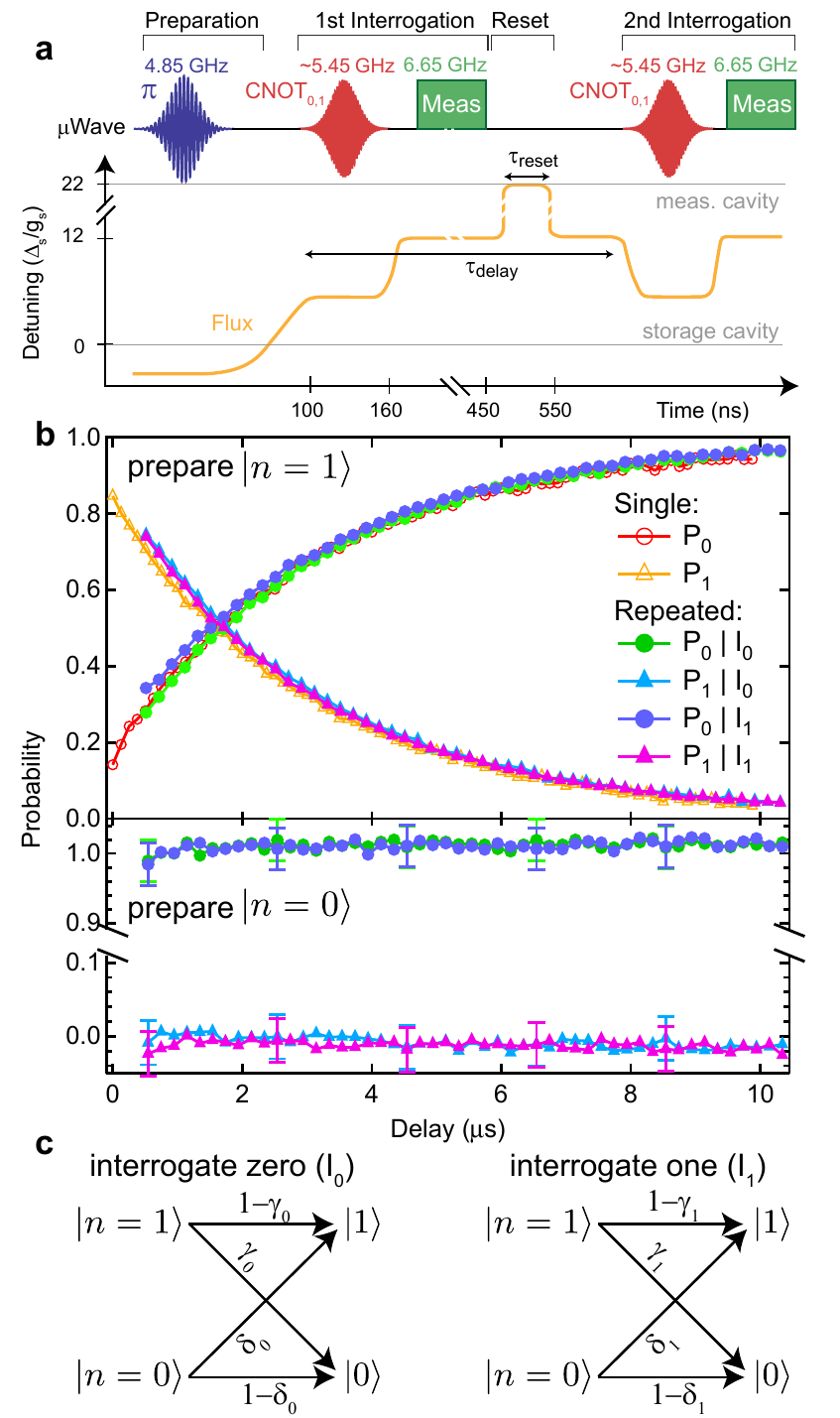}
	\caption{\label{figure 4}Repeated measurements of photons. \textbf{a}, Experiment protocol. A microwave pulse and adiabatic sweep load a single photon into the storage cavity in the preparation step. This photon is interrogated repeatedly by number-selective CNOT gates on the qubit, followed by adiabatic decoupling, qubit readout, and reset. \textbf{b}, Single and repeated interrogation after preparing $\ket{n=1}$ (top) or $\ket{n=0}$ (bottom), ensemble averaged over $\sim 50,000$ iterations. The near-perfect overlap between single and repeated results demonstrate that the protocol is highly QND. \textbf{c}, Transition probability diagrams for the interrogate $n=0$ ($I_0$) and interrogate $n=1$ ($I_1$) processes. We extract $\gamma_0\:(\gamma_1) = 1\:(10) \pm 3\%$ and $\delta_0\:(\delta_1) = 7\:(3) \pm 3\%$.}
\end{figure}

If $\pi$-pulses are used in the interrogation step, measurement results of the average qubit state directly correlate with the probability of being in the states $\ket{n=0}$ or $\ket{n=1}$. Details of the scaling needed to do this transformation when the selectivity is $< 100\%$ are presented in the supplement. These are the desired CNOT operations of the photon meter. If we now insert a variable delay before interrogating, we find that $P_0$  ($P_1$), the probability of being in $\ket{n=0}$ ($\ket{n=1}$) decays exponentially towards $1$ ($0$), as shown by the red (orange) trace in Fig.\,\ref{figure 4}(b). The decay constant of $T_1 \simeq 3.11 \pm 0.02\,\mathrm{\mu s}$ agrees with the linewidth of the storage cavity, $1/\kappa_s = 1/(2\pi\,50\,\mathrm{kHz}) = 3.18\,\mathrm{\mu s}$, measured in a separate, low power ($\bar{n} \sim 1$) reflection experiment.

Strong QND measurements are projective, such that if the measurement observable commutes with the Hamiltonian, the system will remain in an eigenstate of both operators between measurements. Consequently, comparing the results of successive interrogations provides a mechanism to test whether a particular protocol causes additional perturbations on the system. Here, we only compare ensemble average results, because the single-shot qubit readout fidelity for the device is $\sim 55\%$. This is sufficient to reveal processes which change the photon number, and slight technical improvements to interrogation speed or qubit readout fidelity should allow for real-time monitoring of the photon state.

The protocol cannot be repeated immediately, though, because the first interrogation may leave the qubit in the excited state. To circumvent this problem, we use the fast decay rate of the measurement cavity to cause the qubit to spontaneously decay into the 50 Ohm environment. The ``reset" protocol brings the qubit into resonance with the measurement cavity for a time, $\tau_{\mathrm{reset}} = 50\,\mathrm{ns}$, which is sufficient to reset the qubit with probability $\sim 98\%$. The procedure is described in detail in Ref.~\citenum{Reed:2010}.

After resetting the qubit, we can interrogate a second time. The full protocol for a repeated interrogation sequence is shown in Fig.\,\ref{figure 4}(a). The combination of a $\mathrm{CNOT}_0$ ($\mathrm{CNOT}_1$), a qubit measurement, and a qubit reset define an interrogation process $I_0$ ($I_1$). Data for the four possible combinations of interrogating $\ket{n=0}$ and $\ket{n=1}$ are shown in Fig.\,\ref{figure 4}(b) as a function of delay between the first and second interrogations. The data are ensemble averaged over all results from the first interrogation, so we do not observe projection onto number states. Instead, we again observe exponential decay, where the result of the second measurement is essentially indistinguishable from the first, indicating that the interrogation is highly QND.

Deviations from the average measurements of a single interrogation stem from finite photon lifetime in the storage cavity and non-QND processes which cause transitions to other photon numbers [Fig.\,\ref{figure 4}(c)]. Recording the second interrogation results for different delays allows us to subtract the effect of photon $T_1$ and calculate the transition probabilities for the $I_0$ and $I_1$ processes~\cite{Lupascu:2007}. In principle, $I_0$ and $I_1$ can cause transitions to photon numbers outside of the $n\in\{0,1\}$ manifold; however, the absence of statistically significant deviations from $P_0 + P_1 = 1$ suggests that any such effects are negligible. Instead, we observe $\gamma_0\:(\gamma_1) = 1\:(10) \pm 3\%$ and $\delta_0\:(\delta_1) = 7\:(3) \pm 3\%$, demonstrating that this protocol is highly QND.

The protocol presented here is a fast and highly QND measurement of single photons, which we believe can be extended to detect higher photon numbers. It should be possible to demonstrate the projective nature of the interrogation and create highly non-classical states of light via post selection, and eventually with higher fidelity readout it should be possible to observe quantum jumps of light in a circuit.

\begin{acknowledgments}
	We thank Jerry Chow and Michel Devoret for helpful discussions. This work was supported by NSF grants DMR-0603369 and PHY-0653073. J.M.G. was supported by a CIFAR Junior Fellowship, MITACS, MRI and NSERC. L.F. was partially supported by CNR-Istituto di Cibernetica.
\end{acknowledgments}

\section{Supplement}

\subsection{Measured Voltage Scaling}

When the interrogation selectivity is less than 100\%, we need to account for undesired rotations to correctly calculate the state probabilities from the measured voltages. The details of our calibration procedure follow.

If we prepare $\ket{n=0}$ or $\ket{n=1}$ at time $t=0$, when we interrogate at some later time there is an additional probability $p_d$ to decay, giving the density matrices
\begin{eqnarray*}
	\rho^0 & = & \ket{g}\bra{g}\otimes\ket{0}\bra{0},	\\
	\rho^1 & = & \ket{g}\bra{g}\otimes\left\{ p_d\,\ket{0}\bra{0} + (1-p_d)\ket{1}\bra{1}\right\},
\end{eqnarray*}
where $\rho^i$ indicates preparing state $\ket{i}$ at $t=0$. We can model the interrogation pulses as operations which act on $\rho^i$:
\begin{eqnarray*}
	U_0 & = & R_y(\pi)\otimes\ket{0}\bra{0} + R_y(\epsilon')\otimes\ket{1}\bra{1},	\\
	U_1 & = & R_y(\epsilon)\otimes\ket{0}\bra{0} + R_y(\pi)\otimes\ket{1}\bra{1},	\\
	U_I & = & \openone,
\end{eqnarray*}
where $\epsilon$ and $\epsilon'$ are small angles. After interrogation, the integrated homodyne response is
\begin{equation*}
	W_r^n = V_g + \Delta V\cdot \tr(\Pi_e U_r \rho^P U_r),
\end{equation*}
where $n \in\{0,1\}$ is the Fock state of the cavity, $r\in\{0,1,I\}$, $V_g$ ($V_e$) is the voltage measured when the qubit is in $\ket{g}$ ($\ket{e}$), $\Delta V = V_e - V_g$, and $\Pi_e = \ket{e}\bra{e}$.

By abusing the notation slightly and treating $\epsilon, \epsilon'$ as probabilities rather than rotation angles, we can calculate the $W_r^n$
\begin{eqnarray*}
	W_I^0 & = & V_g,	\\
	W_0^0 & = & V_g + \Delta V,	\\
	W_1^0 & = & V_g + \Delta V \cdot \epsilon,	\\
	W_0^1 & = & V_g + \Delta V\left(\epsilon'(1-p_d) + p_d \right),	\\
	W_1^1 & = & V_g + \Delta V\left((1-p_d) + \epsilon p_d \right).
\end{eqnarray*}

We measure these five voltages in calibration experiments and invert the equations to find the parameters $V_g$, $\Delta V$, $p_d$, $\epsilon$, and $\epsilon'$. Note that this does not require perfect preparation fidelity because the model includes decay between preparation and interrogation $p_d$ which will also capture any fixed preparation infidelity. This gives the selectivities, $(1-\epsilon)$ and $(1-\epsilon')$, as well as the preparation fidelity, $(1-p_d)$.

An unknown mixture of $n=0$ and $n=1$ is characterized by a single probability $p$,
\begin{equation*}
	\rho = \ket{g}\bra{g}\otimes\left\{p\,\ket{0}\bra{0} + (1-p)\ket{1}\bra{1}\right\},
\end{equation*}
which produces the responses
\begin{eqnarray*}
	W_0^\rho & = & V_g + \Delta V\left(p + \epsilon'(1-p) \right),	\\
	W_1^\rho & = & V_g + \Delta V\left(\epsilon \cdot p + (1-p) \right).
\end{eqnarray*}
This leads to a simple rescaling to transform $W_0^\rho$ and $W_1^\rho$ into $P_0$ and $P_1$
\begin{eqnarray*}
	P_0 & = & \frac{W_0^\rho - (V_g + \Delta V \epsilon')}{\Delta V(1-\epsilon')},	\\
	P_1 & = & \frac{W_1^\rho - (V_g + \Delta V \epsilon)}{\Delta V(1-\epsilon)}.
\end{eqnarray*}

\subsection{Error Estimate}

The primary challenge in these experiments is obtaining sufficiently accurate and precise control of the qubit frequency to do high-fidelity operations. The narrow bandwidth pulses used in the CNOT operations means that even a few MHz error in frequency control results in a significant rotation error. We use deconvolution techniques similar to those described in the supplement of Ref.~\citenum{Hofheinz:2009}; however, the flux bias current response function drifts on a time scale of about one day, making it difficult to eliminate all classical control errors. Even after applying corrections, there is a remaining spread of $2-3\,\mathrm{MHz}$ in the qubit frequencies over the various realizations of preparation to interrogation delay. This translates into a $2-3\%$ error in the probability to find the qubit in $\ket{e}$ after applying a conditional $\pi$-pulse. The errors bars reported in the lower panel of Fig.\,4(b) are due to this systematic error.

\bibliography{references}
\end{document}